\RequirePackage{fix-cm}
\documentclass[smallextended]{svjour3}       
\smartqed  
\usepackage{graphicx}
\usepackage{amsmath}
\usepackage{amssymb}
\usepackage{braket}
\newcommand{\R}{\mathbb{R}}
\newcommand{\C}{\mathbb{C}}

\newcommand{\ketpsi}{\ket{\psi}}
\newcommand{\kz}{\ket{0}}
\newcommand{\ko}{\ket{1}}

\newcommand{\bellzz}{\frac{\ket{00} + \ket{11}}{\sqrt{2}}}

\begin{document}

\title{Ethical Quantum Computing
}
\subtitle{A Roadmap}


\author{         \
}

\authorrunning{Short form of author list} 

\institute{E. Perrier \at
              Center for Quantum Software and Information  \\
              UTS, Sydney NSW 2000\\
              ORCID: 0000-0002-6052-6798\\
         \email{elija.t.perrier@student.uts.edu.au} }          

\date{Received: date / Accepted: date}

\maketitle

\begin{abstract}
Quantum information technologies, covering quantum computing, quantum communication and quantum sensing, are among the most significant technologies to emerge in recent decades, offering the promise of paradigm-shifting computational capacity with significant ethical consequences. On a technical level, the unique features of quantum information processing have consequences for the imposition of fairness and ethical constraints on computation. Despite its significance, little if no structured research has been undertaken into the ethical implications of such quantum technologies. In this paper, we fill this gap in the literature by presenting a roadmap for ethical quantum computing (and quantum information processing more generally) that sets out prospective research programmes. We summarise the key elements of quantum information processing (focusing on quantum computation) relevant to ethical analysis and set-out taxonomies for use by researchers considering the ethics of quantum technologies. In particular, we demonstrate how the unique features of quantum information processing gives rise to distinct ethical consequences (including in the context of machine learning). We situate quantum ethics at the cross-disciplinary intersection of quantum information science, technology ethics and moral philosophy to assess the impacts of this newly emerging technology. We provide specific examples of how the emergence of quantum technologies gives rise to normative and distributional ethical challenges.  Finally, we set out prospective research directions to help inaugurate the cross-disciplinary field of the ethics of quantum computing.

\end{abstract}
\section*{Declarations}
\textbf{Funding} The authors did not receive support from any organization for the submitted work.\\
\textbf{Conflicts of interest/Competing interests} The authors have no relevant financial or non-financial interests to disclose. The authors have no conflicts of interest to declare that are relevant to the content of this article. All authors certify that they have no affiliations with or involvement in any organization or entity with any financial interest or non-financial interest in the subject matter or materials discussed in this manuscript. The authors have no financial or proprietary interests in any material discussed in this article.\\
\textbf{Availability of data and material} Not applicable.\\
\textbf{Code availability} Not applicable.

\newpage
\section{Introduction}
\label{intro}
Quantum information processing (including quantum computing, quantum communication and quantum sensing) and machine learning represent two of the most significant fields of computational science to have emerged over the last half century. Over the last decade, attention has in particular turned to the ethical implications of machine learning technology, resulting in the emergence of a burgeoning multidisciplinary field dedicated to the ethics of emergent technologies. For example, the ethics of automation and artificial intelligence covers diverse fields, including fair machine learning \cite{caton_fairness_2020,dwork_fairness_2012}, distributive justice, representational justice and jurisprudence. While ethical analyses are advanced or advancing rapidly in many such disciplines, one  omission is the absence of a dedicated research programme into the ethics of quantum information processing technologies more generally. Some recent scholarship has sought to address this gap, including work on high-level legal frameworks for quantum technology \cite{kop_establishing_2021,kop_quantum_2021,johnson_governance_2018}, overview articles \cite{kop_ethics_2021} and responsible quantum innovation \cite{ten_holter_reading_2021} along with earlier high-level work covering societal impacts \cite{Coenen_Grunwald_2017,Coenen_Grinbaum_Grunwald_Milburn_Vermaas_2022,deWolf_2017,DiVincenzo_2017,Grinbaum_2017,Vermaas_2017}. More recently, the World Economic Forum released its \textit{Quantum Computing Governance Principles} \cite{noauthor_quantum_2022}\footnote{Disclaimer: the WEF principles were in part coauthored by this paper's authors drawing on concepts from early drafts of this paper.} that set out a more substantive set of guidelines for thinking about the ethics of quantum information processing technologies, which we denote `quantum ethics'. As we detail in Section 2 below, our paper is comparatively unique in that (i) it proposes a structured roadmap of quantum ethics research while (ii) providing a set of formal taxonomies that connect quantum formalism and ethical theories (and the ways in which unique features of quantum technologies distinguish them from their classical counterpart in ethically relevant ways) in ways we hope are pragmatically useful for researchers.  Such a work which both sets out a comprehensive roadmap for quantum ethics and also connects technicalities of quantum information processing with formalism in ethics has to date been missing from the literature. Thus this work is as much a call to research as it is an attempt to exemplify the type of ethical research into quantum ethics which we believe will be of benefit in theoretical and applied settings. 

The scope of our paper is primarily focused on quantum computation technologies (which we for brevity refer to as `quantum technologies'), though we also touch upon quantum communication (which can in principle be considered forms of computation, if that term is broadly construed to capture information processing itself). While quantum sensing (covering use of quantum systems for measurement or metrology) is traditionally incorporated within quantum information processing, we do not focus on such technologies in this paper.
\subsection{Why quantum ethics?}
\label{subsec:1}
It is legitimate to ask whether the ethical or moral considerations applicable to quantum technologies are significantly different from those relevant to computational technologies in general. To motivate research into the ethics of quantum technology, we must ask: is there anything distinct about the ethics of \textit{quantum} computing in particular? Or can the types of ethical analyses applicable to other technologies simply be transferred to the case of quantum technologies? Is quantum ethics well-motivated? Our answer to this, is yes: while the types of normative considerations applicable to quantum technologies will also apply to other technologies, such as automated algorithmic technologies in general, there are distinct features of both (a) how computing is undertaken in quantum systems and (b) the consequences of quantum computing which distinguish research programmes into quantum ethics. 

At the technical level, the distinct (by contrast with classical computing) characteristics of quantum computation, including its inherently probabilistic nature, the availability of superposition states and resources such as entanglement have distinct implications for the technical implementation of computational ethical regimes reliant on classical features of computation, including proposals regarding the implementation algorithmic governance, fair machine learning, cryptography and representational justice. These technical characteristics of quantum technologies are the basis upon which such technologies offer the promise of an exponential increase in computing power, offering the potential to not only solve certain intractable problems in higher-order complexity classes that are unsolvable on any classical computer, but also to render feasible computations which, while theoretically tractable on a classical computer, may in fact take such time or inordinate resources that they are effectively intractable, at least by time-scales that matter to humans. While much is made of the ability of quantum computers to solve certain exponentially hard problems (such as via the application of Shor's algorithm in cryptographic contexts, a technological feat with profound consequences for geopolitics, privacy and social communication), the more immediate implications of quantum computing may arise through its effective expansion of resources available for classical computation i.e. in enabling classically infeasible computations that directly impact populations (such as simulations of population behaviour) to now be undertaken. Nevertheless, as we discuss below, widespread access to quantum algorithms which may enable both decryption of classically encrypted data and encryption of data in a way that cannot be decrypted (such as by law enforcement) raises serious ethical and moral dilemmas which may, according to some ethical stances, necessitate controls or restrictions upon access to such technology in a way not currently envisaged by stakeholders. 
\\
\\
To this end, we set out a roadmap for ethical quantum computing. Our aim is to contribute to the inauguration of this important and we believe imminently significant field of multidisciplinary research by situating the emergence of quantum computation and its consequences within the broad nexus of ethics, moral philosophy, economics and discourses on ethical computation and algorithms. We aim to provide an overview of key technical features of quantum technologies of relevance to ethical approaches and theories. Of course any such list cannot be comprehensive given the multidimensional nature of quantum computing and quantum information processing as a discipline. \\
\\
We hope to provide researchers across the multiplicity of relevant domains, including quantum science, computer science, ethics, philosophy, social sciences, economics and policy a way to understand the ethics of quantum computing in a cross-disciplinary way, providing a point of entry of researchers in one field, such as quantum science, to another, such as algorithmic ethics and moral philosophy. By doing so, we hope to outline not just why ethical quantum computing is necessarily a multi-disciplinary process, but why understanding and, to the extent possible, managing or even solving ethical dilemmas arising from the use of quantum technologies is necessarily a cross-disciplinary endeavour, where no one field has a monopoly on normative conclusions and where collaboration across the disciplinary aisle is essential.

\subsection{Results and Contributions}
\label{subsec:2}
In concrete terms, the contributions of our paper are as follows:
\begin{enumerate}
    \item A overview and roadmap of research into the ethical considerations and consequences of quantum technologies.
    \item A taxonomy for quantum ethics, detailing the unique characteristics of quantum information processing give rise to distinct ethical consequences for how ethics of quantum computing is characterised, measured and audited (e.g. via quantum fair machine learning).
    \item A mapping of potential research programmes into the ethics of quantum simulations, privacy in quantum contexts and networked quantum systems.
    \item Situating quantum technologies within existing cross-disciplinary ethical frameworks within moral philosophy, economics and computer science.
\end{enumerate}

\subsection{Structure}
\label{subsec:3}
The structure of our paper is as follows. In Section \ref{sec:roadmap}, we set out a high-level proposal for a sequential roadmap of quantum ethics' research covering foundational, technical and governance topics together with features of theoretical and empirical programme design. The remaining parts of the paper then explore specific elements of quantum computing and ethics of relevance to particular limbs of the roadmap.  Section \ref{sec:qcoverview} of the paper provides a short overview of the key technical characteristics of quantum computation relevant to quantum ethics' research and the programmes laid out in the previous section, providing a short synopsis of quantum postulates, quantum information processing and related fields including quantum cryptography. The Section focuses in particular on key differences between classical and quantum computing in order to provide a basis for understanding how the unique nature of quantum computation potentially impacts the use and governance of computational systems in a way that differs from current (and often well-established) approaches in the classical sphere. In Section \ref{sec:qethics} we introduce key concepts in ethical computational science, including algorithmic consequentialism and deontology and frame these within the context of quantum science. Section \ref{sec:qfml} examines how the technical differences of quantum computing, as distinct from classical, lead to differences in algorithmic and computational framings, solutions and require consideration of ethical principles dependent upon the structure of computation itself (rather than just the outcome). The focus of this part is on how approaches to fairness in computation and machine learning vary or are impacted by the use of quantum data or quantum algorithms.

Section \ref{sec:qprivacy} considers individualised ethics and issues arising from the cryptographic consequences of quantum computing, networked quantum computers and the rapidly evolving plans for a quantum internet. In Section \ref{sec:qdistributive}, we discuss a number of economic and distributive aspects of quantum computing, including geopolitical, economic and social welfare (access to resources) questions. Section \ref{sec:conclusion} sets out a summary of our proposal for a roadmap for the development of quantum ethics, offering recommendations for priority research areas.

\section{Quantum Ethics Roadmap}
\label{sec:roadmap}

\subsection{Quantum ethics' literature}
Work on the impacts of quantum technology has been somewhat piecemeal without systematic exposition of (i) how and why unique features of quantum information processing may give rise to unique ethical concerns and (ii) how such research into quantum ethics should be systematically undertaken or the types of taxonomies of use in such research. In particular, scant attention has been paid thus far to synthesising formal technical features of quantum information processing with formal elements of ethical theories (such as establishing concepts of duties, responsibilities, rights etc and then articulating how quantum technologies may affect or alter such normative claims or how ethical constraints might be encoded within quantum information processing protocols themselves). Early attempts to address quantum ethics tended to examine potential societal impacts of quantum technology development primarily from high-level policy perspectives (such as work by Vermaas \cite{Vermaas_2017}) or brief canvassing of potential quantum technological impacts (e.g. DiVincenzo \cite{DiVincenzo_2017}) but with no engagement with technical literature of either ethics or quantum information processing. Other work of a more technical nature by M{\"o}ller and Vuik examined the impact on scientific computing arising from the expansion of computational frontiers on computation that quantum technologies potentially afford \cite{Moller_Vuik_2017} yet offered little in the way of systematic methodology or consideration of the ethical dimensions of such technological developments. Somewhat more germane work on the impacts of quantum information processing was offered by de Wolf \cite{deWolf_2017}, where three key impacts of quantum computing on cryptography, search and simulation were examined. The primary concern raised in that paper regarded the potential distributional impacts of such technologies, but this was only addressed in a cursory way with no engagement in formal distributional ethics (or related work in economics) and as such lacked methodological utility beyond merely raising the prospect of unequal distribution of resources (something not unique or specific to quantum technology per se).  Other papers, such as work by Coenen and Grunwald \cite{Coenen_Grunwald_2017} and latterly Ten Holter et al. \cite{ten_holter_reading_2021} have set out analyses of responsible research innovation practices for quantum technologies. The former of these works largely provides a high-level analysis of where research and governance policy relating to quantum technological development fits within the broader field of technology policy more generally together with an discursive analysis of the framing and narratives of quantum technology by comparison with nanotechnologies. The result of that work is a prescription for research programmes focused on (i) avoiding technological hype, (ii) taking care with respect to narrative framing of quantum technological development. While the discursive treatment of quantum technology is an important feature of the ethical governance of such technology, its conclusions are generic (i.e. they could apply to any technology and indeed are mostly drawn from applying experiences in nanotechnology discourse to quantum), rather than offering anything that may be regarded as a systematic research programme, engagement with formal ethical theories or pragmatic guidance for how one conducts ethical calculus in a quantum setting. The work in \cite{ten_holter_reading_2021} is of more use in that it sets out recommendations for how to (empirically) engage stakeholders across the quantum sector when developing responsible research practices in the area. Related work has also considered how concepts and technological impacts from quantum information processing may be communicated pedagogically and to broader audiences (e.g. Grinbaum \cite{Grinbaum_2017}), yet only tangentially touches on how such discursive reframings may or may not give relate to ethical questions.  

More recent work on the ethical implications of quantum technology has focused on governance protocols and situating quantum ethics within technology ethics more broadly. An example is Kop's work on situating quantum ethics within broader principles applicable to related technology (such as artificial intelligence) \cite{kop_establishing_2021} which has seen an explosion of interest and governance. Other work in the governance context \cite{kop_quantum_2021} examined impacts upon intellectual property principles of quantum technology which, while not directly related to ethics, would indirectly affect ethical matters of concern, such as property rights and distributional/access issues surrounding quantum technology. Other work by Johnson \cite{johnson_governance_2018} provided a useful overview of how `soft governance' instruments, such as frameworks, principles and the like may be of use for quantum governance. While recent and earlier attempts at exploring quantum ethics are important in canvassing a number of issues, they have generally lacked any systematic consideration of how and where quantum ethics as a field should evolve. They also mostly lack engagement with formal normative theories (e.g. little mention of consequentialist versus deontology for example), the expansive literature on technology governance and jurisprudence (with the exception of \cite{johnson_governance_2018}) or risk-management literature (which is of particular relevance to how one operationalises ethical constraints). All such works lack any consideration of or engagement with formal computational science approaches on encoding ethical criteria within the functioning of technology itself (as we discuss below). Thus it is clear that as a cross-disciplinary field, the ethics of quantum information technologies has thus far lacked any cohesive framework for how research into the area should be conducted and any conceptual exploration of how ethical constraints would or could actually be encoded within the design of quantum technologies themselves.

\subsection{Roadmap}
As an emerging cross-disciplinary field at the intersection of a raft of expansive fields, quantum ethics as a discipline faces considerable definitional challenges from its inception. Any research programme seeking cover under the flag of such polysemous terminology must provide, if it is to develop with theoretical rigour and practical utility, a well-defined and bounded set of concepts and methods. In this paper, we do not lay claim to any canonical formulation of quantum ethics. But we do seek to provide introductory structure, concepts and methods to such a field by setting out a strategic roadmap and guideposts for quantum ethics research programmes (\textbf{QEP}). The aim of the roadmap is to propose for researchers and stakeholders in quantum technology, governance and ethics a set of research imperatives that will facilitate scholarship and thinking around the ethics and governance of quantum computation.

The roadmap is set out in a loosely sequential manner. We argue that for QEP to contribute meaningfully to debates over quantum ethics and regulation, firstly what we denote the `technical foundations' of such programmes must be fleshed out. Such an initial research programme would be driven by ethical design principles and therefore necessitate an understanding of the ways in which quantum technologies technically differ from their classical counterparts with respect to properties such as system control, information processing and output when it comes to imposing ethical constraints upon such systems. Placing this programme at the beginning, as it were, then enables researchers to situate quantum technologies within the expansive landscape of technological ethics and governance literature in order to facilitate the development of the second part of the roadmap, research into quantum governance models. With a solid understanding of the similarities and differences of quantum technology with respect to ethical design and control, QEP can then proceed to analysing the extent that (a) existing governance regimes are suitable for managing quantum technologies or (b) new or novel governance or institutional frameworks are needed to manage quantum technologies. For example, quantum communications technology may or may not give rise to new considerations outside the current remit of communications governance. A third part of our roadmap then proposes research into risk-management techniques specifically tailored for quantum technologies and the roles, duties and responsibilities, guided by existing risk theory. The objective of such research is to provide more granular and useful research for the diverse array of quantum stakeholders involved in the technology's development and use, such as governments, corporations, academia and the public and private sectors. We summarise our roadmap taxonomy in Table 1 and elucidate details of such programmes below.

\begin{center}
\begin{figure}
\begin{tabular}{ |p{0.03\textwidth} |p{0.2\textwidth}||p{0.7\textwidth}|  }
 \hline
 \multicolumn{3}{|c|}{Quantum Ethics' Roadmap} \\
 \hline
 1. & Technical Foundations  & \textit{Foundational theory}:  
 \begin{itemize}
     \item formalising ethical constraints in terms of quantum information processing mechanisms
     \item devising theoretical models to represent ethical classification of quantum procedures and outcomes 
 \end{itemize}
 
 \\
  & & \textit{Ethical quantum design}: 
\begin{itemize}
    \item design of quantum systems or algorithms to implement ethical or moral constraints
    \item design of interfaces with quantum technologies in order to satisfy moral or ethical criteria 
\end{itemize}
  
  \\
  \hline
  2. & Ethical Governance & \textit{Quantum governance}: 
  \begin{itemize}
      \item research into applicability of existing technological governance frameworks to quantum technologies
      \item development of new institutional and governance practices to cater for unique ethical or moral consequences of quantum computation
  \end{itemize}
  \\
   & & \textit{Quantum stakeholders}:
   \begin{itemize}
       \item research into ethical duties, obligations and responsibilities of different stakeholders in quantum sectors (e.g. governments, corporations)
   \end{itemize}
   \\
   \hline
   3.& Ethical Risk Management & \textit{Quantum risk protocols}: 
   \begin{itemize}
       \item research into upside and downside risks of quantum computing and related technologies 
       \item research regarding adaptation of existing risk protocols and standards for quantum technologies and institutions
   \end{itemize}
   \\
   & & \textit{Quantum transformation management}:
   \begin{itemize}
       \item research into likely transformational impacts of quantum technologies on different constituencies
       \item research regarding optimal ways to manage such transformational impacts balancing opportunities and risks
   \end{itemize}
   \\
 \hline
\end{tabular}
\caption{A roadmap setting out topics for quantum ethics' research programmes.}
\end{figure}
\end{center}

\subsubsection{Contextualising quantum ethics}
Any QEP undertaking should situate itself within and draw upon existing approaches to technological ethics. It is important therefore that each limb of any QEP clearly identify the extent to which (i) existing theory or practice can be applied to research into and solving ethical dilemmas arising from quantum technologies; and (ii) novel approaches to theory, practice or governance are required. The world abounds with technology ethics' theories, frameworks, practices and norms of institutional governance. QEP should where possible avoid `reinventing the wheel' as it were and instead draw upon existing approaches . For example, the rich body of national and international jurisprudence regarding the governance of technology provides a ready-made basis upon which governance practices for quantum stakeholders and institutions may be based. Conversely, it is important that research into quantum ethics does not simply treat quantum technologies as equivalent to other computational or engineering technologies where to do so would be inappropriate. As we argue below, there are unique characteristics of quantum computation, quantum information processing and certain quantum technologies which motivate the development of quantum-specific ethical epistemology. 

QEP are necessarily cross-disciplinary. And as with any cross-disciplinary project, a considerable part of any QEP will involve a process of translation between concepts in one field, such as theoretical quantum computation, to others, such as ethics. At a pragmatic level, this will involve, as is common within quantum-related subdisciplines, extracting `quantum equivalents' of classical results, such as in quantum communications theory or indeed quantum computing itself (witness the Deutsch-Church-Turing thesis as one example). Thus a fruitful line of research for researchers interested in QEP is to consider how to map the epistemology of ethics to an applied quantum context. Quantum computation has the benefit of being able to draw upon extensive fields at the cross-roads of computer science, ethics and decision-theory. 

Moreover, when seeking to undertake cross-disciplinary work in computational ethics, it is incumbent on researchers to be aware that the field of ethics itself contains a rich set of decision-procedures and resources for reasoning through ethical dilemmas. A myriad of disciplines from formal ethics, to moral philosophy, to economic sciences to name a few provide well-established techniques for systematically defining and resolving (or at least going some way to resolving) such dilemmas. So too do jurisprudential disciplines which, for example, provide often canonical examples of applied deontological reasoning, from the establishment of duties to important concepts of responsibility, standards of care and liability.

\subsection{Technical Foundations}

The first element in a QEP roadmap involves establishing the technical basis of quantum ethics. Such a \textit{technical foundations} branch would consist of laying out the foundational theorems and framing of quantum computational ethics.

\subsubsection{Foundational Theory}
The first part of QEP requires the translation of concepts from ethics into a formalism applicable to quantum computation. As we argue below, a number of the unique features of quantum, as distinct from classical, computation give rises to differences of ethical consequence. For example, the nature of quantum computation differs from classical computation in significant ways, such as: (i) in how information is encoded in quantum states (such as in superposition states and via basis and phase encoding); (ii) in how quantum systems evolve in the presence of absence of noise; (iii) in the inherent ontological stochasticity of quantum systems (as distinct from their classical counterpart); (iv) in how quantum systems may be measured, controlled or audited; (v) in the in the availability of certain resources such as entanglement which in turn give rise to novel information architectures unavailable in the classical setting; and (vi) in the unique consequences arising from the quantum nature of such computation, including the capacity to undertake computations that are inefficient or even intractable on classical computers. 

Each of these distinguishing features potentially carries with it consequences for ethical analysis of quantum computation and quantum systems generally. This is most apparent when the ethical imperatives in question relate directly to the manner in which computation is undertaken itself. A motivating example which we discuss in some detail below is the case requirements for algorithms to meet certain ethical fairness criteria or the proscriptions on the inclusion or combination of certain features (such as protected attributes on the basis of race or ethnicity). Classically, compliance with ethical requirements, such as those imposed by anti-discrimination principles or even legislation, can take the form of constraints upon inputs (e.g. certain inputs are proscribed), constraints upon form of a program and upon how computation unfolds (for example, prohibitions on combining sensitive confidentialised datasets in ways that allows individuals to be identified - a requirement typical when using Census microdata provided for academic research by government agencies), or constraints upon outputs, such as ensuring classification criteria (such as statistical parity or Lipschitz fairness) are met within certain tolerances. In a quantum setting, in which quantum algorithms are utilised instead of (or in conjunction with) classical algorithms, the ways in which such ethical criteria are met will differ: the form of quantum computation is governed by Hamiltonian formalism; a quantum algorithm is stochastic, outcomes occur probabilistically rather than deterministically (for example, running the same program over and over again will not necessarily lead to identical output); the manner in which the computation may be controlled also differs; and overall there is much less certainty and because in order to undertake quantum computation, interactions with the quantum system must be kept to a minimum (e.g. only via measurements or control) in order to preserve quantum coherences, information about the exact nature of the execution of a quantum algorithm and its true evolution will usually remain. We expand on foundational details in Section \ref{sec:qcoverview} (providing an overview of distinguishing features of quantum as distinct from classical computation) and Section \ref{sec:qethics} (covering ethical framing of computation itself).

\subsubsection{Ethical quantum design principles}
An important initial focus of our roadmap and any QEP is to connect with the literature on `ethical design', which considers pragmatic ways in which ethical constraints or criteria may be incorporated into the design of systems, institutions and technology themselves \cite{hoven_handbook_2015,hoven_design_2005}. Examples in ethical research include the study of how institutional design mechanics can facilitate the realisation of particular normative imperatives (e.g. Rawlsian principles of justice), encoding moral principles into system mechanics. As ethical design research notes, top-down, deductive, approaches where ideal moral or ethical actions are chosen and systems are then engineered in response suffer from a lack of specificity to be actionable, leading variously to exploration of reflective equilibrium or action-guiding methods. Moreover, given the ubiquity and increasing role of technology as the primary medium for human interactions (such as via social media or  designed environments), ethical imperatives require a consideration of the design interfaces between technologies and their users. Specific research questions include (i) technical specifications regarding how to encode ethical criteria within quantum algorithms (such as phase or basis encoding within Hamiltonians); (ii) research into different techniques for the control of quantum systems affects the degree of control that can be exerted on quantum algorithms to ensure ethical criteria are met; and understanding how the inherently stochastic nature of quantum computation affects. Sections \ref{sec:qethics} and \ref{sec:qfml} in particular provide specific technical examples of how the design of quantum algorithms could be shaped by ethical considerations, drawing on, among other things, examples in classical computational planning literature. 

The challenge of any ethical reasoning is to justify \textit{why} a particular ethical constraint, theory or practice should be chosen out of many (to the extent distinct). While to remain useful, QEP should restrain consideration of such meta-ethical questions, these are in part unavoidable and are not confined to quantum or any other technologies. For example, the types of concepts of fairness or equality, the types of rights or meaning of rights, according to which quantum systems should be managed is itself usually going to be contested: the values embodied in one set of ethical principles may not be acceptable to other constituencies or appropriate in other contexts. The challenge in part with computational technologies is that they in some sense `put to proof' implicit assumptions or requirements about \textit{how}, for example, an ethical decision-procedure should unfold because of the necessity to encode and design for that procedure within algorithms themselves. That is, the formal requirements of algorithms which must be designed using logic can force a clarity or form upon ethical discussion which can reveal the inherent logic (or vagueness) in an ethical position. Rudimentary examples from the world of classical machine learning, for example, include results demonstrating the intuitively obvious fact that ethics usually involves trade-offs, such as the intractability of sub-group statistical parity for arbitrary groups \cite{kearns_empirical_2019}. Ethical design must also contend with the appropriate way to reason through ethical or moral values when considering how to functionally or operationally embed them within quantum systems.

\subsection{Quantum Governance and Responsibility}

Together with more philosophical or formal theoretical ethical considerations, pragmatic considerations regarding the governance and regulation of quantum technologies represent a key feature of QEP. While still nascent, there is now an emerging literature that canvasses the ways in which existing governance methodologies would apply to prospective quantum technologies and ways in which new or novel governance approaches may be needed as a result of the unique aspects of quantum systems (as discussed below). 

\subsubsection{Quantum Governance}
The quantum governance branch of QEP should aim to provide research into the jurisprudential and governance. As with any technology, governance is a two-way street: quantum technologies, their design, use and distribution, must as with any technology be subject to the rules and conventions. Similarly, laws, regulations and ethical frameworks must be tailored to and fit for purpose with respect to quantum technologies. This may mean, for example, updating laws governing the use, combining or distribution of data containing protected or sensitive information in a way that takes into account the unique characteristics of quantum computation (to combine information into superposition states, or limitations on accessing quantum information) directly within legislative regimes. Thus a useful line of research for QEP concerns an auditing and accounting of the extent to which technology ethics and jurisprudence is fit for the distinct features of quantum technologies. Recent work in this direction aims to provide a high-level legal framework that proposes certain ethical design principles \cite{kop_establishing_2021}. Other work considers the ways in which intellectual property jurisprudence, such as copyright, patent and trademark regulation, would apply to quantum technologies \cite{kop_quantum_2021}. As with other fields of technology ethics, QEP can inform how to integrate various ethical considerations into competing objectives related to quantum technology. For example, while much technology ethics is concerned with downside risks, there are also costs and risks from overly constraining or shackling the development of quantum technologies in the name of moral virtue. Thus understanding the technical trade-offs or compatibility of mooted moral values with the need to foster innovation in quantum technologies is an example where QEP can add value to debates over governance.

\subsubsection{Quantum Stakeholders}
A cornerstone of many fields of ethics, especially deontological and jurisprudential ethical subdisciplines, considers ethics from the perspective of rights, duties and obligations. In an applied and pragmatic setting, the regulation and governance of quantum technologies will necessarily involve conforming quantum governance regimes into existing jurisprudential frameworks which ascribe rights, duties and responsibilities upon a variety of persons, including individuals, associations, governments and corporations. As discussed in ethical design literature, the design of institutions that will not only govern quantum technologies, but also be involved in their development and deployment should factor in the particular ethical and moral outcomes that are sought. As such, an important feature in QEP focused on quantum governance will involve an initial mapping of `quantum stakeholders', those constituencies, institutions and agents involved in quantum technological ecosystems. Such stakeholders include governments, multinational institutions, academia, corporations and individuals, the rights, duties and responsibilities of which in relation to quantum technology will differ depending on context:
\begin{enumerate}
    \item \textit{Governments and multinational institutions}. The regulatory, policy and economic power of states makes them the primary stakeholders in any quantum governance regime both at the intranational level (via municipal laws and regulations) and internationally, via treaties, concordats and other multilateral and bilateral instruments. Governments have also been to date the primary economic investors and drivers in the development of quantum technologies and quantum computation, such as via direct programmes and, in particular, national defence institutions (such as DARPA). QEP programmes could focus upon a fine-grained ethical analysis of the types of duties governments would owe and other normative responsibilities towards their constituencies and how these in turn would shape governmental policy and legislative responses. Furthermore, given the tight nexus between investment in quantum technology and defence, research programmes in this area may also focus on the geo-strategic implications of quantum technologies. Indeed that the majority of funding for quantum outside academia has to date has largely been via defence-connected institutions and the inclusion of quantum in national strategic outlooks (e.g. China's five-year plans), such analysis is both timely and relevant to real-world considerations. A similar analysis would also apply, albeit in a different jurisprudential context, regarding multilateral institutions. 
    \item \textit{Academia}: the majority of research and technical expertise in quantum technology has traditionally been situated (and remains) within academic institutions such as universities. Indeed quantum computation as a discipline emerged out of largely theoretical considerations brought on by the emergence of quantum systems and limitations facing researchers seeking to simulate real-world systems (such as Feynman). Furthermore, public funding of academic quantum programmes represents a considerable benefit. Thus  consider QEP programmes in this regard may consider the types of ethical obligations upon academic institutions to maintain and promote the development of quantum science, but also to meet other moral or ethical constraints. Similarly, QEP can usefully be directed towards guides or even heuristics for the types of ethical considerations that may be relevant to other research programmes. For example, cryptographic research with the potential to impact privacy would be one such area. Indeed, recent proposed constraints upon the export of quantum technologies in the name of national security evidence the importance of establishing ethical and governance protocols for quantum research from the outset.  
    \item \textit{Other stakeholders}. Other stakeholders of interest include private corporations (users and developers), together with individual users and developers, each having roles in the functioning and development of the quantum sector. From an ethics' perspective, QEP programmes can usefully draw in extensive literature on individual, collective and group deontological ethics in order to reason through the different duties and responsibilities of such stakeholders. For example, what duties do private sector stakeholders involved in the development of quantum technologies have regarding access, controls or maintaining standard with regard to such technology? How might the ethical considerations for a quantum algorithm designer working on cryptography differ from a quantum engineer working in a laboratory, or political decision-maker considering quantum policy? 
\end{enumerate}
An early example of proposed quantum governance principles directed towards stakeholder interests that attempts to incorporate values-based ethical design is provided by the World Economic Forum's recent \textit{Quantum Computing Governance Principles} \cite{noauthor_quantum_2022}. The principles set out framework based upon core ethical values of collective well-being, accountability, inclusiveness, equity, non-maleficence, accessibility and transparency. These in turn form the basis of different themes and topics germane to the ethics of quantum computing, such as development goals, opportunities and risks. The principles aim to provide a taxonomic approach of use to different quantum stakeholders in their deliberations and responses and represent an example, albeit generalised, of the type of frameworks that more in-depth QEP research could build upon.

\subsection{Quantum Transformation and Risk Management}
Another useful source of research for practical use in both quantum governance and technical approaches to quantum ethics involves the application of well-established risk assurance and management frameworks to quantum technologies. As with all technologies, understanding risks is central to making pragmatic decisions regarding the use of technology and in informing the types of ethical or moral constraints that are deemed acceptable. As we discuss below, quantum computing comes with the promise of extraordinary paradigm-shifts in a number of areas, while also representing a potentially profound disruption to core features of modern integrated societies via its impact on cryptographic and communications protocols. QEP research applying methods derived from international standards and risk management protocols such as ISO 31000 \cite{purdy_iso_2010} in the development of quantum-specific risk management tools would be of considerable use for stakeholders, such as governments and corporations, familiar with such risk management settings. Research programmes in this area would adapt familiar risk management paradigms in terms that are pragmatically relevant for particular quantum stakeholders as risks will depend on the interests, duties and obligations of particular stakeholder groups which vary. For example, governments and government agencies bear a different set of ethical duties and obligations to private corporations or academia involved in quantum technology development. As such, the ways in which those institutions manage risk in general and quantum-specific risk will likely differ and can be informed by research programmes tailored towards a quantum context. 

More broadly, if the potential of quantum technologies is realised, they are likely to have extraordinary and profound transformational impacts. The heightened computational promise of full fault-tolerant quantum computing has the potential to transform sectors and industries, such as pharmaceuticals, health, medicine and industrial technology. As we discuss further in Section \ref{sec:qdistributive}, the capacity to simulate and model complex systems, solving certain intractable currently inefficiently-solvable problems in finance, logistics and other areas all potentially could impact different constituencies and stakeholders. A major branch of QEP would then include identification upside and downside risks from these technologies and the likely institutions and populations they may impact. In practical terms, this may involve ethics' research into (a) how to measure such risks or (b) institutional frameworks for responsibility and change management within organisations affected by quantum technology impacts. Further detail of this type of risk assessment process is specified in recent work on quantum governance \cite{noauthor_quantum_2022}, covering ethical considerations around access to quantum technology, innovation, raising awareness, cybersecurity, privacy and sustainability.





\section{Quantum Computing - An Overview}
\label{sec:qcoverview}
\subsection{Overview}
Quantum computing is a vast multi-faceted discipline, spanning specialisations across mathematics, theoretical and applied computer science, condensed matter physics, quantum field theory, communications and information theory and cryptography. In this section, we outline the key technical characteristics \textit{quantum} computation of relevance when assessing the ethical impact of quantum technologies. We summarise the elementary features of quantum computation relevant for understanding how, for example, quantum algorithms differ in key respects from their classical counterparts. Understanding such differences, we argue, is important for research programmes assessing the comparative similarities and differences between the ethics of quantum and classical technologies. 
\subsection{Postulates}
Quantum information processing is characterised by constraints upon how information is represented and processed. Here we briefly summarise the postulates of quantum mechanics as applicable to quantum computing, following \cite{nielsen_quantum_2011}. This section is a truncated synopsis of a summary in [\textbf{Redacted reference (2)}] and is included for the benefit of researchers who may not be familiar with the main tenets of quantum computing. We present the following synopsis in comparative fashion against similar or analogous concepts in classical computer science.

\subsubsection{State space}
Classical computation is typically represented using stateful paradigms in which a system is represented by a \textit{state}, typically comprising a set of properties represented as an element (vector) of a characterising the system in that state. Computation is then presented as transitions among states as per classical Turing machine or finite-state formalism (see \cite{sipser_introduction_2012} for an introduction). Such transitions may be either deterministic or non-deterministic and are usually represented by a transition operator or stochastic transition matrix which sets out the probabilities of the system transitioning from one state to another under the operation of some computational process. While the specific transition may not be determinable in advance (e.g. it may be probabilistic) and while the exact information about the classical state may not be practically accessible (due to, for example, limitations in hardware or overly complex system descriptions), \textit{in principle} a classical computational system subsists in a definite \textit{single} state: it is \textit{ontologically} determined. Uncertainty about what state a classical system is in is thus not \textit{ontological}, but epistemological, a representation of the uncertainty of our knowledge rather than any intrinsic indeterminacy about the system state itself. This underlying \textit{in principle} determinacy of classical computational states underpins much of the design thinking and assumptions behind the use, control and auditing of classical computational systems \cite{burghes_control_2004}.

Quantum computation, by contrast, is a fundamentally distinct computational paradigm, though contains within it classical computation as a \textit{limiting} case where ontologically the probability of being in a state is unity (not unlike how classical field theories within physics are considered limiting cases of more general quantum field theories). In a foundational sense, the apparent indeterminacy of quantum states can be either epistemological, as a result of incompleteness of state information (a mixed state, see below) or fundamentally ontological such that the state is informationally complete (a pure state). The unique ontology field of quantum foundations is addressed primarily within the field of quantum foundations \cite{harrigan_einstein_2010,bartlett_reference_2007} where debates about the ontological(ly distinct) status of quantum systems is at issue, for example, in debates over hidden variables' models (where uncertainty about quantum states is due to epistemological incompleteness or inaccessibility of variables or features of quantum systems which exist but cannot be measured) and certain seminal results (such as Bell's celebrated theorems \cite{bell_einstein_1964}).\\ 

Classical computation's unit of analysis is the bit, a binary value $x \in \{0,1\}$ which, usually in sequences of bits, characterises a state. Information is sequentially encoded in classical bits, for example via bit-strings $x_1,...,x_n \dot{=} (x_n) \in \{ 0,1\}^n$. The bit-string may only subsist in any one evaluated state at any time $t$. The evolution of classical systems is then represented by mappings (or Boolean functions defined over) subsets of $\{ 0,1\}^n$.  Quantum systems are, by contrast, described by more structure owing to the fact that they must obey the laws of quantum mechanics. By comparison with the classical case, quantum systems are completely described by (unit) state vectors within a complex (rather than real) valued vector space called a Hilbert space $\mathcal{H}$, a vector space over $\C$ equipped with an inner product. As we explicate below, this additional structure is necessitated by the ways in which quantum systems evolve and quantum states interact, primarily in order to preserve certain symmetry properties and statistical probability measures. The most basic formulation of quantum computation involves analysis of the states and evolution of two-level quantum systems, expressed as elements of two-dimensional Hilbert spaces, named qubits. A qubit $\ketpsi$ within such a Hilbert space with basis $\{\kz,\ko\}$ is represented as:
    \begin{align}
        \ketpsi = a\kz + b\ko
        \label{eqn:qubit}
    \end{align}
Qubits share in common with bits being represented via the binary-valued basis states. However, they differ in ontologically and formally in fundamental ways from classical bits. Firstly, as seen in (\ref{eqn:qubit}) above, it can be seen that unlike classical bits which are $0$ or $1$, the qubit state is a linear composition of \textit{both} basis states. Secondly, each basis state is accompanied by coefficients $a,b \in \C$, denoted as amplitudes. These amplitudes combine in ways that enable a sequence of amplitudes to be represented a complex-valued vector space mentioned above (e.g. $(a,b) \in V(\C)$). These amplitudes reflect the probabilities of measuring $\ketpsi$ to be in state $\kz$ or $\ko$. Here, $\ketpsi$ is in state $\kz$ with probability $|a|^2$ and $\ko$ with probability $|b|^2$. Thus quantum states are characterised by the probability distributions associated with measuring specific outcomes. This rule is known within the field as the Born rule which fundamentally maps specific properties of quantum states (amplitudes) to the probability of measuring that state as such. 

We explicate measurement protocols in more detail below, but it is important to be aware that colloquial phrases such as `the system is in state' or `the qubit is in state' are really terms of art referring to inferences drawn from the measurement statistics that arise (and are conditional upon) repeated measurements of identically prepared initial states. Already, this compositional linearity fundamentally distinguishes quantum from classical states with implications for how information is encoded in such systems, how that information evolves (or is processed) and how such systems may be controlled: all key elements, as we articulate, in any attempt to control or conform quantum systems to particular rules, ethical or otherwise. 

Furthermore, unlike the classical case, in which information may only be encoded in the value taken on by a bit, information in qubits may be encoded in two ways: (i) via basis encoding (similar to bits); but also (ii) by relative phase encoding, that is, via encoding the amplitudes $a,b$ in ways that represent information. The amplitudes $a,b$ are subject to further constraints, known as normalisation constraints, imposed in order to ensure the probability of all measurement outcomes sums to unity (i.e. to guarantee and appropriate probability measure on the space). This is denoted formally via the condition upon unit vectors that $\braket{\psi | \psi} = 1$ (that is $|a|^2 + |b|^2=1)$. 
Here $\braket{\psi | \psi'}$ denotes the inner product of quantum states $\ketpsi,\ket{\psi'}$. The inner product has a rich formal theory associated with it, but intuitively it can be thought of as the degree of `overlap' between two states and thus is used as a measure of their similarity or difference. All classical bits can indeed be represented as being in either the basis state $\kz$ or $\ko$ with unit probability for either e.g. $|a|^2=1$ such that $\ketpsi=a\kz$. Note, for completeness, that while such a state would be indistinguishable in terms of measurement outcome from a classical bit with value 0, technically speaking the amplitude $a \in \C$ can take any root of unity in $\C$, so need not be the real number 1 (whereas classically this must be the case). It should be noted that quantum information processing literature distinguishes between physical qubits, which is ostensibly any physical system that exhibits or meets the criteria for being a qubit, and logical qubits, which may include a whole system of physical qubits whose collective state behaves according to qubit criteria. 

Quantum systems are also highly sensitive to noise and distortions which can give rise to errors in quantum states (e.g. changing the energy level or valuation from $\kz$ to $\ko$ or interfering with coherences). As a result, the imperative of quantum system design is to achieve what is known as \textit{fault tolerant} quantum computation, a technical term with related theorems that refers to the idea that if physical (qubit) error rates can be reduced below a certain threshold, then logical (qubit) error rates can be reduced arbitrarily. Various theoretical and technical means of error correction, such as encoding quantum information in certain codes which are robust to noise or can self-correct up to acceptable probabilities, are a major focus of quantum computing research and distinguish the field from its classical counterpart, where error correction is relevant but less foundational (owing to the robustness of physical classical computing devices).

For completeness (for readers who may delve more deeply into quantum formalism), more advanced treatments of quantum computing (and quantum information processing) tend to utilise what is known as density operator formalism, which is ostensibly a useful way of representing probability distribution over which a quantum state $\ketpsi$ subsists.  In density operator formalism, a quantum system is described via the positive density operator $\rho$ with trace unity acting on the state space of the system. If the system is in state $\rho_i$ with probability $p_i$ then $\rho = \sum_i p_i \rho_i$. Density operators are represented usually via matrices.   

\subsubsection{Encoding information}
There are four primary methods of encoding data in quantum systems \cite{schuld_supervised_2018} (a) basis encoding (e.g. encoding a bit-string $(001)$ as qubit basis states $\ket{001}$), (b) amplitude encoding, where information is encoded in the amplitudes of quantum states, together with (c) qsample encoding and (d) dynamic encoding (which we omit). Typically, encoded data is then placed into a superposition state in order to leverage quantum advantages as a result. While considered generally an unremarkable step, there are potential ethical issues to be explored about the ethics of encoding data (such as from a fair representation learning \cite{mcnamara_provably_2017}). For example, does placing data into superpositions potentially enable the combining (via quantum interference) of protected attributes into new features in a way that is ethically proscribed (e.g. combining information about individuals to create features responsible for unfair computational outcomes)? While speculative, such questions of representational justice are important within certain fields of ethics and jurisprudence.
    
\subsubsection{Evolution}
A key feature of any programme that seeks to address the technical foundations of quantum ethics is an appreciation of the distinct ways in which quantum computation evolves by comparison with its classical counterpart. Quantum systems are constrained in how they may evolve in order to maintain the uniquely quantum features of the system (such as coherences) that distinguish them from classical systems. For example, quantum state transitions must conform to specific rules designed to preserve probability measures. Quantum systems are also modelled variously in terms of close and open systems. Closed systems are toy models which treat a quantum system as self-contained and which disregard the effects of various noise sources or other interaction effects from anything outside the system (denoted the environment). What counts as the system or environment will depend upon context. For example, in typical closed-system quantum control toy models which disregard distortion or noise effects, apparatus by which a quantum system is controlled may be conceived as part of the system as a whole (represented by a set of control Hamiltonians), despite being physically distinct from the physical qubits themselves. Open quantum systems are models that take into account the presence of noise which can interfere with the quantum characteristics of a system (such as coherences or energy of the system) in a way that can give rise to errors or the irreversible loss of information. 
Closed quantum systems (which we focus on in this paper for simplicity) evolve over time $\Delta t= t_1 - t_0$ via unitary transformations: 

\begin{align}
    U=\mathcal{T}_+ \exp\left(-i\hbar \int_{t_0}^{t_1}H(t) dt \right)
\end{align}
Such unitaries represent solutions to the time-dependent Schr{\"o}dinger equation of motion governing evolution:
    \begin{align}
        i\hbar \frac{d\ket{\psi(t)}}{dt}=H(t)\ket{\psi(t)}
    \end{align} 
    where $\hbar$ is set to unity for convenience and $H$ represents the linear Hermitian operator (Hamiltonian) of the closed system. The $\mathcal{T}_+$ term is the time-ordering operator (a requirement of quantum field theory). The dynamics of the quantum system are completely described by the Hamiltonian operator acting on the state $\ketpsi$ such that $\ket{\psi(t)} = U(t)\ket{\psi(t=0)}$. In density operator notation, this is represented as $\rho(t_1) = U \rho(t_0) U^\dagger$. Typically solving the continuous form of the Schr{\"o}dinger equation is intractable or infeasible, so a discretised approximation represented as a discrete quantum circuit (where each gate $U_i$ is run for a sufficiently small time-step $\Delta$t) is used (and the impact of $\mathcal{T}_+$ is assumed away).
 
    The Hamiltonian $H$ of a system is the most important tool for mathematically characterising dynamics of a system, encoding the computational processing of data encoded into quantum states and specifying how the quantum computation may be controlled. To the extent that computational processes or outcomes are important to ethical criteria (for example, understanding the dynamics giving rise to bias or discriminatory outcomes, or ensuring fair representation learning in a quantum setting), they must usually be encoded in the Hamiltonian of quantum systems. Modifications to computational subroutines in quantum computing must be done indirectly, via adjustments to the Hamiltonian which steer quantum systems towards target states using unitary evolution. 
\subsubsection{Measurement}
Information about or within quantum systems (which characterises those quantum systems or object information encoded within) is extracted by quantum measurements and its associated formalism. Quantum measurements are represented as sets of measurement operators $\{ M_m\}$. By operating on a system with a measurement operator, the apparatus , where $m$ indexes the outcome of a measurement (e.g. an energy level or state indicator). The probability $p(m)$ of outcome $m$ upon measuring $\ketpsi$ is represented by such operators acting on the state such that $p(m) = \braket{\psi | M_m^\dagger M_m |\psi}$ (alternatively, $p(m) = \text{tr}(M_m^\dagger M_m \rho)$ with the post-measurement state $\ket{\psi'}$ given by: 
    \begin{align}
        \ket{\psi'} = \frac{M_m \ketpsi}{\sqrt{\braket{\psi | M_m^\dagger M_m | \psi}}}
        \label{eqn:postmeasurementstate}
    \end{align}
    The set of measurement operators satisfy $\sum_m M_m^\dagger M_m = I$, reflecting the probabilistic nature of measurement outcomes which are reflected in distributions over those outcomes. Density operator (matrix) formalism $\rho$ in effect provides a way to represent probability distributions for quantum measurement (and thus state characterisation) in the language of operator algebra. In advanced treatments, POVM formalism more fully describes the measurement statistics and post-measurement state of the system. In this case, we define a set of positive operators $\{  E_m \}=\{M^\dagger_m M_m\}$ satisfying $\sum_m E_m=I$ in a way that gives us a complete set of positive operators (such formalism being more general than simply relying on projection operators). As we are interested in probability distributions rather than individual probabilities from a single measurement, we calculate the probability distribution over outcomes via Born rule using the trace $p(E_i) = \text{tr}(E_i \rho)$. This formulation is important to quantum fairness metrics discussed below.

\subsubsection{Composite systems}
State vectors $\ketpsi$ in the Hilbert space $\mathcal{H}$ may be composite systems, described as the tensor product of the component physical systems $\ket{\psi_i}$, that is $\rho_i = \sum_j p_{ij}\ket{\psi_{ij}}\bra{\psi_{ij}}$. We also mention here the importance of open quantum systems where a total system comprises a closed quantum system $H_S$ and environment $H_E$ ($H = H_S + H_E + H_I$) where the last term denotes an interaction term, which is typically how noise is modelled in quantum contexts. 
%
%
Other key concepts necessary to understand the formalism below include: (a) \textit{relative phase}, that for a qubit system, where amplitudes $a$ and $b$ differ by a relative phase if $a = \exp(i\theta)b, \theta \in \R$ (as discussed below, classical information is typically encoded in both basis states e.g. $\kz,\ko$ and in relative phases); (b) \textit{entanglement}, certain composite states (known as EPR or Bell states), may be entangled. For example, for a two-qubit state: \begin{equation}
    \ketpsi = \bellzz
\end{equation}
measurement of $0$ on the first qubit necessarily means that the second qubit is also in the state $\kz$. Entangled states cannot be written as tensor products of their component states i.e. $\ketpsi \neq \ket{\psi_1}\ket{\psi_2}$. Such states are crucial to important emerging quantum domains of quantum communication and quantum cryptography; (c) \textit{expectation}, expectation values of an operator $A$ (e.g. a measurement) can be written as $E(A) = \text{tr}(\rho A)$; (d) \textit{mixed} and \textit{pure} states, quantum systems whose states are exactly known to be $\ketpsi$, i.e. where $\psi = \ketpsi\bra{\psi}$ are pure states, while where there is (epistemic) uncertainty about which state the system is in, we denote it as a 'mixed state' i.e. $\rho = \sum_i p_i \rho_i$ where $\text{tr}(\rho^2)< 1$ (as all $p_i < 1$); (e) \textit{commutativity}, where two measurements are performed on a system, the outcome will be order-dependent if they do not commute, that is, if $[A,B]\neq 0$; and (f) \textit{no cloning}, unlike classical data, quantum data cannot be copied (for to do so requires measurement which collapses the state destroying the coherent superpositions that encode information in amplitudes). 
\\
\\
We omit a universe of other characteristics of relevance to quantum ethics, including error-correcting codes (encoding mechanisms designed to limit or self-correct errors to achieve fault tolerant quantum computing) which, while relevant, are beyond the scope of this paper. The above postulates and characteristics of quantum computing also provide the basis for taxonomic distinctions between (a)  \textit{quantum data} and \textit{classical data} (e.g. different state characterisation) and (b) \textit{quantum information processing} versus \textit{classical information processing} (by the requirement, for example, of any coherence-preserving quantum process to respect unitary evolution). 
In the following sections, we explore the direct and indirect ethical consequences of distinct characteristics of quantum computing identified above.

\section{Quantum Ethics and Ethical Algorithms}
\label{sec:qethics}
\subsection{Overview}
An important characteristic of prospective ethical quantum computing programmes is being able to informatively situate quantum ethics research within the broader fields of ethical inquiry. Research into the ethics of algorithms, computation and technology is a vast enterprise, covering classical ethics, philosophy and computer science, together with emergent cross-disciplinary fields such as fair machine learning, computational population ethics and research into regulatory platforms. In this section, we sketch a number of key landmarks and concepts of relevance to quantum ethics' research, with the aim of identifying foundational questions in quantum ethics. Put another way, while specialisation in quantum ethics is necessary and inevitable, it is incumbent upon researchers in quantum ethics to be sufficiently aware and engage with existing ethics' literature. To this end, we follow a proposed framework articulated in [\textbf{Redacted reference (1)}].

\subsection{Ethical computation in quantum contexts} 
What it means for computation (as distinct from the consequences of computation) to be ethical or normatively satisfactory requires an appreciation of the nature of computation itself. This is true also of quantum computing. The classification of a quantum computation \textit{itself} as ethical or not can usefully be understood according to the following taxonomy of algorithmic computation. We adopt a distinction between the process of computing and its outcomes (though note such a distinction can be reframed in terms of either using typical models of computing, such as via Turing machine formalism). While such a distinction oversimplifies the complexities of normative ethics, it is useful as a first framing for researchers. 

\subsubsection{Ethical quantum computation}
A quantum computation may be ethical if it consists of (i) \textit{ethical procedures}, a form of algorithmic deontology (concerned with what computational evolutions or pathways are or are not permissible) where the process of computation itself (unitary evolution) satisfies some ethical constraint or requirement or (ii) \textit{ethical outcomes}, a form of algorithmic consequentialism. Each classification depends upon the underlying normative criteria specifying what counts as ethical. Ethically classifying the former requires understanding how system Hamiltonians (which govern unitary evolution) encode ethically-relevant computations for example (for example, precluding consideration of protected attributes from the action of Hamiltonians). Assessing ethical outcomes also requires understanding quantum measurement statistics, as the ability to identify or audit an evolving quantum system is ultimately contingent upon probabilistic measurement outcomes (we cannot `open the black box' of a quantum computer, as it were, to audit its operations in the same way as may be in principle available when using a classical computer). An additional consideration arising from the indirect way in which quantum states and processes are identified is the accuracy with which quantum states and processes must be identified for classification as ethical. For example, state characterisation is often performed via tomographic methods, which seek to reconstruct characteristics of quantum states from measurements \cite{greenbaum_introduction_2015,dariano_quantum_2001}. However, the resources required to tomographically characterise quantum systems scales exponentially, rendering full characterisation of quantum systems largely infeasible. As such, researchers seeking to ethically classify quantum states or processes must reconcile with inherent uncertainty and vagueness about the actual states/processes of quantum systems. Reckoning the consequences of such informational incompleteness is not unique to quantum information processing (such incompleteness is a feature when seeking to characterise many classical systems). It does, however, speak to the inherently probabilistic nature of quantum computing which has consequences for the degree of certainty or confidence required for ethical classification of quantum computations.
\subsubsection{Modelling ethical procedures}
The ethical characterisation of quantum computation can draw fruitfully upon the extensive literature on constraint satisfaction problems and planning within computational science and control theory \cite{kolobov_theory_2012}. Such approaches represent a form of design-thinking for ethics (discussed in Section \ref{sec:roadmap} above) in computation that aims to encode ethical behaviour in the functioning of algorithmic and autonomous systems. In planning and control literature, one is concerned with how to model and execute planning procedures. Such classical MDP models have more recently been applied to consider \textit{ethical compliance} of sequential decision procedures with specific ethical criteria. For example, Svegliato et al. \cite{svegliato_ethically_2021} propose an MDP-based model for ethically compliant autonomous systems that aims to decouple ethical compliance from task completion. In this model, decision-procedures are subject to moral principles within ethical contexts. Such concepts are formalised within an MDP framework following a typical Bellman-style \cite{bellman_dynamic_1956} decision-procedure described by the tuple $\braket{S,A,T,R,d}$. The system subsists in a finite state $s \in S$, with an action (decision) $a \in A$ giving rise to a transition among states $s_t \to s_{t+1}$. The probability of transitioning is given by $T$ and is conditional in $p(s'|s,a)$, with $r \in R$ a reward function and $d$ a parameter modelling which initial state $s_0$ the system starts in. Solutions to MDPs are policies which are probability distributions over actions $\pi: S \to A$ representing the probability of taking action $a$ given state $s$. The unfolding of computation is then modelled stochastically in terms of such stochastic state-transitions, where optimising consists then of finding an appropriate policy $\pi^*$ that maximises an expected discounted cumulative reward satisfying a typical Bellman optimality equation. In this model, moral principles (say utilitarianism) form constraints upon the particular policies with optimal policies being those that maximise the dual objectives of conformity with the moral principle while maximising the overall objective. Such an approach might be described as a `stateful' approach to ethical computation where satisfying ethical constraints on how a computation should unfold is equivalent to proscribing states that the computational system should not subsist within. Of course, not every ethical consequence of computational technology necessarily can or should be framed in such a stateful manner e.g. the distributional ethics of quantum technologies is a distinct question from whether a particular computation transitions to a proscribed state or not. But such an approach is (i) very much consistent with the well-established literature on encoding constraints within computational architecture itself \cite{mouaddib_handling_2015} and as a result (ii) useful for thinking about cases where ethical or normative requirements mandate that computations are not carried out in a particular way or do not result in particular outcomes.

In these sorts of models, some choice of ethical standard (e.g. a principle of utilitarianism) becomes encoded in the decision-procedures of such algorithms. Moral and ethical principles manifest in proscriptions or prescriptions upon state transitions and their probabilities. For example, proscribed states are then modelled as `dead ends', e.g. a moral constraint that requires a zero probability of transitioning to a proscribed state \cite{mouaddib_handling_2015}. Thus ethical constraints can be equivalently modelled as either (a) constraints on the types of states $s \in S$ into which a system may transition or (b) constraints upon the transition functions among states. As the broad field of computational planning attests, the challenge lies in issues such as whether a system is fully \textit{controllable} and the extent to which full information about the system is known or knowable, indeed whether a particular end-state is reachable given an arbitrary or unknown initial state $s_0$ (for example, it may be technically impossible to reduce the transition to an ethically proscribed `dead end' state to zero no matter what the policy is) \cite{kolobov_theory_2012}. Classical Markov decision procedures (MDP) and control theory provide a well-studied framework within which to model the ethical classification of quantum computational processes. 

For example, in the quantum setting, imposing moral principles as constraints on the types of states that a system may subsist within and transition to manifests formally as constraints within the Hamiltonian $H$ of the quantum system. To the extent the system is controllable, then the imposition of moral or ethical constraints in terms of state transitions would also necessitate specific control architecture, which in turn would be represented via, for example, control terms within the Hamiltonian. For example, in certain ethical frameworks, certain judgments by a decision-procedure system may be proscribed: a car running a red light, a classification algorithm breaching certain fairness criteria (see below). Because decisions by such a system are modelled in terms of the \textit{state} that the system is in, then to meet ethical policy criteria (e.g. as set out in \cite{svegliato_ethically_2021}), one would in a quantum setting require terms in the Hamiltonian that prohibit the quantum system from evolving into that proscribed state $\ket{\psi'}$. Whether a state evolves to such a proscribed state is then a function of the Hamiltonian $H$ as the Hamiltonian ultimately controls the transition probabilities among states (formally expressed through propagators and the Dyson equation). However, what counts is whether such a system can be controlled or influenced. This depends on the initial state of the quantum system $\ket{\psi_0}$ and how it evolves according to $\ket{\psi(t)} = U(t) \ket{\psi_0}$ where $U(t)$ is as described in Section \ref{sec:qcoverview}. In control theory, this is modelled by partitioning the Hamiltonian into controllable $H_c$ and uncontrollable $H_d$ (`drift') terms:
\begin{align}
    H = H_c + H_d
\end{align}
Controlling a quantum system, or at least rendering it less likely to transition to proscribed states then becomes a question of how to fashion the control Hamiltonian $H_c$ which is effectively the way in which a system is by design controlled to maintain constraints. The probability of then transitioning to a particular state can be represented, analogously with the transition matrix $T$, via a density matrix $\rho$ which acts in effect as a representation of a probability distribution over states. The exact form of a control Hamiltonian required to meet a particular ethical constraint will depend on context, quantum algorithm design and a raft of other factors. Nevertheless, the foregoing provides an example of the type of encoding of ethical constraints in quantum systems of relevance to technical foundations QEP. 

The approaches described above are by no means the only models according to which quantum algorithms may be classified as ethical, but they provide a guide (particularly starting from a classical perspective) as to how ethical classification and control may work in a quantum context and an indication of the type of research topics that the technical foundations part of a quantum ethics' roadmap could entail.

\subsubsection{Auditing and provability of ethical quantum computations.} 
Methods of auditing or proving that a quantum computation is ethically satisfactory (and that ethical constraints are being complied with) are important potential areas of quantum ethics' research as they are for classical systems. Auditing of quantum computation poses distinct challenges compared with the classical case (see \cite{kearns_empirical_2019}) because of the computation cannot be measured directly (colloquially, one cannot `open the black box'), but rather is dependent upon reconstructing information about the computation via repeated measurements on a quantum system. Attempts to measure `in train' as it were can cause quantum states to collapse (as per (\ref{eqn:postmeasurementstate})) or disturb quantum systems in a way that interferes with information about the system in the first place. Secondly, research into auditing of quantum algorithms involves consideration of proof techniques for quantum computation. The provability of quantum computations \cite{vidick_quantum_2016} requires consideration of different strategies and resources (such as for provers and verifiers, including witness states or entanglement \cite{ji_mipre_2020}), together with quantum correctness logics \cite{feng_quantum_2020}. 

\subsection{Four C's of ethical computation}
In addition to ethical classification and auditing/provability considerations, the specific nature of quantum computing has implications for (following [\textbf{Redacted reference (1)}]) computability, complexity, consistency and controllability. We briefly sketch the significance of these categories for questions about the ethics of quantum computing below.

\subsubsection{Computability and complexity} 
Computability questions for the ethics of quantum computing require two questions to be asked: (i) is the proposed ethically constrained quantum computation computable? That is: is it even a type of computation that is possible for a quantum computer?; and (ii) is the computation efficient or ethically feasible given resource constraints? For example, imposing particular ethical constraints on the unitary evolution of an algorithm may increase the resources required to undertake the computation. Relatedly, as discussed below, ethical constraints can sometimes be considered as restricting the subspaces $\mathcal{H}_i$ of $\mathcal{H}$ within which $\ketpsi$ is permitted to lie or evolve within. Other considerations we group within computability include the degree of certainty or precision required of a quantum computation, that is, do ethical criteria impose unrealistic levels of precision on measurement or state reconstruction which would render such ethical constraints on the computation infeasible? Additionally, the complexity classes of certain quantum computations (see \cite{aaronson_quantum_2013}) is an important characteristic for assessing their feasibility. Indeed some authors regard complexity as in practice a more relevant property of algorithmic architecture than computability per see (see Lloyd \cite{lloyd_turing_2012} for example). This is because the computation may be outside the bounds of what is tractable for even a quantum computer or, more practically, the complexity specifications help give an indication of the resources required to actually undertake the computation in an ethically constrained manner.
\subsubsection{Consistency and controllability} 
Consistency and controllability of quantum computations are two more important features that impact on their ethical framing. The types of considerations relevant here are (i) \textit{process (deontological) consistency}, whether similar decisions ought to follow similar methodologies; and (ii) \textit{outcome (consequentialism) consistency}, whether it is normatively permissible for two different quantum algorithms (or even one algorithm) to come to different output classifications given the same input data (e.g. quantum analogues of Lipschitz conditional fairness see [\textbf{Redacted reference (2)}]). Of the Four C's, \textit{controllability} is among the most studied. Controllability asks whether an ethical algorithm is sufficiently controllable, both in terms of steering it towards desired end-states, but also the extent to which unitary evolution is itself controllable (e.g. via Hamiltonian controls). Quantum control (see \cite{dalessandro_introduction_2007} for an introduction), namely controlling quantum computations to satisfy ethical criteria is therefore an  important feature of research programmes examining technicalities of quantum ethics.

Other issues to consider when analysing the ethics of quantum computations themselves cover those applicable to ethical algorithmic analysis generally, including the appropriateness of (inevitable in most cases) quantum computational heuristics and the types of ethical trade-offs that particular quantum algorithmic architectures presage. 

\section{Quantum Fair Machine Learning}
\label{sec:qfml}
\subsection{Overview}
In this section, we provide examples of how the distinctly quantum nature of quantum computation impacts technical results in ethical computational science. To illustrate the way in which the different nature of quantum (as distinct from classical) computation affects technicalities of ethical research programmes, we focus on how the use of quantum data and quantum algorithms, including quantum machine learning-based algorithms, changes how ethical constraints can or would be imposed on those algorithms. In concrete terms, we summarise how various technical results in the field of fair machine learning (the leading technical discipline at the interface of computer science and ethics), such as parity measures, metrics and auditing, is altered when quantum information is involved. Our synopsis here draws on work by the authors of this paper [\textbf{Redacted reference (2)}] which presents in more detail quantum analogues of canonical results in fair machine learning (such as the use of amplitude amplification to achieve statistical parity and Lipschitz fairness conditions in quantum algorithmic contexts). 

\subsection{Quantum and Classical Fairness}
Fair machine learning is a sub-branch of computer science concerned with how classical algorithmic systems can be assessed against ethical criteria (such as bias or discrimination in data or outcomes) and strategies for imposing ethical standards (such as fair outcomes or representational justice) on such algorithms \cite{caton_fairness_2020,chouldechova_fair_2017,del_barrio_review_2020,dwork_fairness_2012}. Quantum fair machine learning (QFML) [\textbf{Redacted reference (2)}] is a prospective cross-disciplinary area of applied quantum computing where quantum or hybrid (classical-quantum) algorithms (ranging from traditional quantum algorithms to those designed according to machine learning principles) are subject to fairness constraints at the preprocessing, in-processing (modelling) or post-processing (post-measurement) stage. In this sense, the proposal for QFML represents a more technical approach for analysing how the unique features of quantum computation interplay with ethical requirements. Our exposition focuses on the use of quantum information processing for solving ethically constrained optimisation problems involving classical data encoded into quantum systems (in the $CQ$ quadrant as per \cite{aimeur_machine_2006}).

\subsubsection{Parity measures} 
Fair machine learning is characterised by the designation of some \textit{fairness criteria} according to which (usually the outcomes of) algorithms are adjudged. There are a multitude of fairness criteria (see \cite{caton_fairness_2020} \cite{del_barrio_review_2020} for examples. A simple example is \textit{statistical parity} which demands fair (equiprobability of) classification by a classifier $Y=1$ of two or more subgroups ($g_1,g_2$) of a population i.e. $Pr(Y=1|g_1)=Pr(Y=1|g_2)$.  As discussed in the proposal for QFML [\textbf{Redacted reference (2)}], parity in a quantum context can be framed in the following way: fairness criteria effectively partition the Hilbert space into direct sums $\mathcal{H} = \oplus_i \mathcal{H}_i$ such that satisfying fairness criteria is realised via the equiprobability of $\ketpsi$ residing in an arbitrary subspace $\ket{\psi_i} = \{ \ketpsi | \ketpsi \in \mathcal{H}_i\}$. The fairness criterion is then expressed with respect to a suitably chosen (POVM) measurement operator $M_i$ that partitions $\mathcal{H}$ into subspaces $\mathcal{H}_i$ as (quantum fairness):
    \begin{align}
    \braket{\psi_i | M_i^\dagger M_i | \psi_i} &= \braket{\psi_j | M_j^\dagger M_j | \psi_j}\\
    \text{tr}(\rho_i M_i) &= \text{tr}(\rho_j M_j)
 \end{align}
where $\rho_i$ corresponds to $\ket{\psi_i}$. That is, the probability of measuring the state $\ketpsi$ in $\mathcal{H}_i$ is equal for each subspace. In addition, distance metrics on quantum computers differ in ways that lead to modification of classical fairness constraints. One example (see [\textbf{Redacted reference (2)}]) is for Lipschitz conditioned fairness (where similar individuals normatively should be subject to similar outcomes), where classical states metrics are replaceable by density operators and quantum metrics, such as trace distance and fidelity, as ways of assessing state similarity.  
\subsubsection{Mitigating unfairness}
Classical techniques for mitigating unfairness have mixed transferability to  quantum computational settings. Strategies for ensuring this in a classical context often involve preprocessing data (redacting data, or sampling to adjust training data composition or regularisation). Undertaking quantum computations satisfying such fairness criteria is framed somewhat differently due to the unique aspects of quantum computing discussed above. For example, quantum data and processes cannot be inspected directly, rather one is limited to a probabilistic distribution over measurement outcomes i.e. classification is inherently probabilistic. Moreover, one usually can only sample a portion rather than the entirely of $\mathcal{H}$, thus even the distribution is an estimate. Furthermore, there are no guarantees that quantum states (upon which fairness measures depend) are reliably distinguishable in general. One remedial approach is to leverage quantum algorithms to satisfy fairness constraints. An example of this is presented in [\textbf{Redacted reference (2)}], where the application of Grover's amplitude amplification algorithm is shown to evolve a quantum system in order to achieve statistical parity (i.e. equiprobability of measuring $\ketpsi \in \mathcal{H}_i$ for $i=1,...,n$) via rotating $\ketpsi$ within $\mathcal{H}$.

The discussion above only touches upon a few of the types of ethical issues that research programmes into quantum fair machine learning. Research directions include (a) developing quantum analogues of classical fair machine learning techniques, (b) exploring prospective QFML methods in the presence of noisy or decohering systems (including, for example, leveraging the dissipative characteristics of such systems to simulate the dissipative characteristics of neural networks \cite{schuld_quest_2014}), (c) a detailed study of how specifically quantum computational resources, such as entanglement, affect fair machine learning techniques and (d) the application of privacy and quantum cryptographic protocols in ethical contexts. In the remainder of this work, we examine normative and ethical considerations arising from quantum computation including for privacy, distributional ethics and geopolitics. 


\section{Privacy and Cryptography}
\label{sec:qprivacy}
\subsection{Overview}
Quantum computing has significant implications for privacy on the technical, ethical and legal fronts. Construed broadly, research programs considering how quantum technologies impact privacy have the benefit of expansive literature on quantum cryptography. Research programmes into ethics of quantum privacy should factor in results from the vast ethical and philosophical work on privacy (see \cite{decew_privacy_2018} for a summary). For example, within the philosophical literature, a distinction is often drawn \cite{schoeman_philosophical_1984} between (a)  \textit{reductionists} who argue that there is no normatively distinct privacy interests per se because all privacy concerns ultimately are derivative from supervening concerns around utility or other types of wrongs (such as distress); and (b) \textit{coherentists}, in which privacy is viewed as a distinct and normatively valid interest. Here, we examine some of the ethical implications for privacy, encryption and information accessibility arising from quantum technologies.

\subsection{Differential privacy and quantum computing}
Differential privacy is a leading proposal within computational science for enabling the public distribution of information while retaining as confidential protected or sensitive classes of information (such as individual information). 
The modern canonical treatment of differential privacy lies in seminal work by Dwork et al. \cite{dwork_differential_2008} who define a randomised algorithm $Q(X_i)$ as satisfying $\epsilon$-differential privacy if, given two datasets $X_1,X_2$ differing by certain sensitive elements, for all subsets $R$ of the image $Q=\text{img}(Q(X_i))$ the following relation is satisfied:
\begin{align}
    Pr(Q(X_1) \in R) \leq \exp(\epsilon) Pr(Q(X_2)\in R)
\end{align}
Intuitively this results means that the ability to distinguish whether a calculation involved the sensitive data (thus being able to estimate how the calculation changed as a result and infer features of the sensitive data) reduces asymptotically to zero in the limit in the presence of sufficient noise. Practically, differential privacy then provides an estimate of the noise or randomness needed to guarantee differential privacy thresholds.  What does privacy mean in a quantum context?
One of the advantages of the inherently probabilistic nature of quantum data and quantum information and the no cloning theorem is that it is much more difficult to extract information such that the problems of maintaining privacy in quantum settings while being able to utilise quantum information converge on a similar point. One candidate for addressing privacy in a quantum context is Aaronson et al.'s work on \textit{gentle measurement}, where parallels are drawn between the need to retain the confidentiality of sensitive information while utilising that information with the need, in a quantum context, to retain quantum coherence in quantum data while being able to perform informative measurements on the quantum system \cite{aaronson_gentle_2019}. Gentle measurement regimes can in principle facilitate the utility of quantum data while retaining quantum coherence and preserving ethically mandated confidentiality. Thus to the extent that normative privacy concerns drive ethical criteria regarding the use of quantum algorithms, existing quantum methodologies themselves may provide means of satisfying or addressing those concerns.

\subsection{Quantum cryptography ethics}
A significant field of quantum computing is the broad field of quantum cryptography, including emerging disciplines such as quantum key distribution and elements of quantum communications. Cryptographic results in quantum information science have provided significant motivation to expansive research into the role and possibilities of encryption in the presence of scalable fault-tolerant quantum computers. Much interest in this area - and quantum computing generally - has been driven by Shor's seminal result on the ability of quantum computers to efficiently factor prime numbers in a way that could, under certain conditions, enable breaking of classically-based encryption protocols \cite{shor_polynomial-time_1997}. That result in itself illustrates the significant ethical and indeed geopolitical consequences of quantum computing. 

While still technically remote, the potential to decrypt classically encrypted communications holds enormous implication for not just individual privacy concerns, but also the confidentiality upon which economic activity is undertaken across networked communication systems. Similarly, the strategic consequences of being able to crack the codes of adversaries has profound strategic implications on the cybersecurity front and for direct armed conflict (as the role of code-breaking during WWII demonstrates). Since those results, extensive work has been undertaken into quantum cryptography with its own set of ethical consequences, these include, for example, the potential for encryption regimes that are effectively non-decryptable even on quantum computers. \\
\\
The availability of technology to enable effectively unbreakable cryptography also raise a host of ethical questions that research programmes into quantum ethics may consider. We sketch a few below.

\subsubsection{Distributional issues} 
In a similar vein to distributional ethics questions about who may access quantum computational resources, the potential of quantum technologies to decrypt and encrypt data can also be framed in terms of population ethics and distributive justice. That is, if access to quantum technologies provides an advantage to some (by way of unbreakable encryption) and disadvantage to others (by rendering their data potentially subject to decryption), then the availability and distribution of cryptographic resources becomes an ethical question. To this end, the types of considerations we explore in the following section (on distributional ethics of quantum computing generally) will apply to the case of cryptographic resources. For example, research may wish to explore whether the distribution and extent of encryption of data can be modelled via social welfare or expected utility means: what are the consequences, for total social welfare, of ubiquitous access to encryption? Should encryption resources be constrained by pricing systems or other regulatory frameworks? 

\subsubsection{Decryption ethics} 
While much privacy research in computational sciences is directed towards the ethical imperatives of protecting privacy from data breaches or decryption, there are many theoretical and practical cases where it may be ethically desirable or imperative to deprivatise or decrypt. For example, normative Rawlsian arguments \cite{bay_ethics_2017} can be made that unbreakable encryption itself unethical, undermining other types of normative requirements and claims regarding social cohesion and ordering.  Thus it is important for research programmes in cryptography and privacy to also factor in how technologies can be used to deprivatise data, break codes or interfere with quantum and classical communications.

\section{Distributive Ethics, Strategy and Quantum Simulations}
\label{sec:qdistributive}
\subsection{Overview}
The distributional ethics of quantum computing resources is an important research programme for quantum ethics. The computational capacities and impact of quantum computing on economic modes of production and communication have the potential for profound structural impact. Quantum computing can properly be construed as an economic resource. The distribution of quantum computing technologies, including who may access them, how access may be mediated (or restricted) and the types of economic (and social) activities for which quantum computers are available is an inherently normative question. Decisions about distribution of quantum technologies involve technical, economic, political and geopolitical questions. For example, establishing networked quantum systems, such as is envisaged via proposals for the quantum internet \cite{wehner_quantum_2018,kimble_quantum_2008,lloyd_infrastructure_2004} will be affected in part by distributional questions about access to such technologies (and even moreso, geopolitical questions). 

In this sense, because the distribution of networked quantum technologies intrinsically affects \textit{both} the technicalities of quantum communication and social welfare, the ethics of the distribution of quantum systems is somewhat distinct from other technologies whose distribution does not impact the actual functioning of the devices (that is, distribution correlates with both functionality and social welfare utility in the case of networked systems). From one perspective, it is understandable that quantum researchers have not turned their mind to the ethics of distributing quantum technologies, in a sense one needs to build the technologies first. However, the normative impact of such technologies does presage the study of their distributional consequences. Research programmes into these questions will require cross-disciplinarity from the beginning. To this end, we outline a number of ways in which well-developed normative and ethical theories (and debates) may usefully inform such research.
\subsection{Distributional ethics for quantum technologies}
In this section, we outline how results and frameworks from subfields of ethics and economics can help guide research into distribution of quantum technologies. For example, certain impossibility results, such as those around well-ordering of preferences population ethics

\subsubsection{Social choice theory and welfare axiology} 
Welfare axiology covers how the ordering of preferences among populations affects the types of distributional systems (such as distribution of economic resources) which are normatively preferable or possible. Conceptually, it is a broad term covering population ethics, the economics of social welfare and public financing, along with social choice and decision-theoretic social sciences. The underlying question of such approaches lies in how to order populations in order maximise some measure of total social welfare, relying upon an ordering of preferences and method of aggregating them (such as a type of expected utility) \cite{list_social_2013}. Work in this area involves considering how the networked distribution, say of individuals, affects welfare aggregation functions, including cluster analysis \cite{kleinberg_inherent_2016} \cite{pierre_barthelemy_median_1981} along with graph-theoretic methods \cite{hammond_consequentialist_1988}. The network-theoretic aspect of both distributional ethics and quantum network engineering provides a basis through which to study the ethics of quantum technological distribution. In concrete terms, the task can be framed as a multi-constraint optimisation problem, where  utility $S(I_i,Q_i,A_{I,Q})$ is presented as a function of individual utilities $I_i$ (of the population) and desirable quantum measures of utility $Q_i$ of quantum network assets (for example, channel fidelities), along with an adjacency matrix $A_{I,Q}$ that encodes a graph detailing connection information among individuals, quantum network assets (where they are located) i.e. $\underset{A}{\max}\, \mathbb{E}(S(I_i,Q_i,A_{I,Q}))$. The output of such research would be an optimal (or more optimal) understanding of how to distribute quantum network technologies to improve social welfare while meeting minimum technical requirements. Moreover, research at the intersection of social choice theory and quantum technology can draw on the emerging field of computational social choice which covers computational paradigms involving social choice mechanisms, such as multi-agent systems \cite{brandt_handbook_2016}.
    
\subsubsection{Impossibility theorems}
Population ethics and social welfare economics is also characterised by in-depth study of impossibility results which ultimately show the impossibility (or inconsistency) of satisfying multiple welfare criteria or obtaining a well-ordering of population ethics satisfying such criteria, including for example optimal distributions of quantum technologies. Examples include seminal work by Arrow  \cite{arrow_difficulty_1950} and Arrhenius \cite{arrhenius_impossibility_2000}. Practically, expected utility approaches must necessarily be bounded by the resource constraints of calculating such expected utilities (see critiques of \cite{mcgee_we_1991,simon_behavioral_1955,klaes_conceptual_2005}). Conversely, it may be that quantum computing does permit the calculation of certain utility functions which would otherwise be infeasible or intractable for classical computing. Put another way, utilising quantum computers to expand the boundaries of bounded rationality presents an interesting set of research programmes (see section in this work on the ethics of quantum simulations and decision-procedures). 

\subsubsection{Quantum Simulations} 
One of the motivations for quantum computing has been the capacity of quantum computers to simulate physical systems in a way which is intractable on classical computers or, even if tractable, is practically infeasible (due to spatial or temporal resource constraints). The promise, at least as speculatively put, is that the scalable and exponential computational power of quantum computing devices will expand the frontiers of the types of problems whose solutions which may be feasibly computed, across domains such as finance, economics, multi-agent systems. Such speculations must be contrasted with theoretical and technical realities of quantum computing, including the limited number of problems for which quantum computers are provably superior and considerable difficulties in engineering fault-tolerant scalable quantum devices \cite{preskill_fault-tolerant_1997}. Nevertheless, to the extent that quantum computing does offer the promise of, for example, simulating classically complex interacting systems, such as dynamic social or economic systems, then the ethics of such quantum simulations should be a matter of research and debate. 

In this section, we canvass the ethical consequences of such enhanced simulation technologies. Two proposed research programme topics in particular are worth mentioning in this context: (a) \textit{individual simulation} that is, to what extent could the scalable resources of full fault-tolerant quantum computing enable highly precise or accurate simulations of individuals, including their preferences, psychology and behaviour? This is not a uniquely quantum question (as classical simulations of individual behaviour have a long history) but does become a question for quantum ethics if it is the unique aspects of quantum computation that render such simulations feasible. 

The ethical consequences of such simulation capacity are of course profound and worth exploring; and (b) \textit{multi-agent simulations}, which similarly asks the question of how quantum computing power may push the frontiers of the simulation of multi-agent systems \cite{drogoul_multi-agent_2002,parsons_game_2002}. For example, the ability to more precisely model larger numbers of interacting agents has potential consequences across a range of social, economic and geostrategic contexts. At a practical level, it entails consequences for modelling the interaction of driverless cars, including driverless cars with distinct ethical subroutines (e.g. two car algorithms may have different conclusions about how to react to an imminent accident) and multi-agent reinforcement learning contexts \cite{barton_measuring_2018}.




\section{Conclusion}
\label{sec:conclusion}
The rapid spread of quantum computing and increasing utility of even near-term devices motivates the establishment of research programmes dedicated to analysing the ethical contexts, questions and consequences of these technologies. In this paper, we have sought to contribute to these ethical imperatives by presenting a taxonomic roadmap addressing the key features of research programmes into quantum ethics for use across disciplines, covering the following:
\begin{enumerate}
    \item \textit{Technical characteristics of quantum systems}. Research programmes in this field require a sufficient understanding of the ways in which quantum computation differs from its classical counterparts.
    \item \textit{Cross-disciplinary and ethically informed}. Research programmes into quantum computing should factor in, from the beginning, an understanding of relevant philosophical and ethical research in related fields such as decision-theory, moral philosophy and social choice theory. Results from such fields can assist understanding the types of ethical constraints, classifications and consequences of relevance and avoid research programmes proceeding with naive or simplistic ethical frames.
    \item \textit{Quantum fair machine learning}. We provided an overview for prospective research programmes as to how fair machine learning, the most computationally technical of existing multidisciplinary fields, can be extended to the quantum realm.
    \item \textit{Privacy and cryptography}. We elucidated a number of ethics' research questions arising from privacy and cryptographic consequences of quantum technologies. Two research directions proposed in particular include into differential privacy and the normative distribution ethics of cryptographic resources.
    \item \textit{Distributive ethics}. We demonstrated how quantum computing and networked quantum communications gives rise to the types of ethical considerations relevant to the fields of population ethics and distributive justice. In doing so, we provided the outlines of issues that research into these areas may consider, including expected utility measures in networked settings and the relevance of impossibility theorems. 
\end{enumerate}
We have also in various parts of this paper aimed to provide worked concrete examples of how ethical research into quantum computing ethics may proceed. As we have argued above, quantum computing and quantum technologies raise a host of profound ethical and moral consequences which motivate the considered development of quantum ethics as a cross-disciplinary field. Our paper we hope will provide a useful set of guideposts for researchers interested in this emergent and important discipline.

\bibliographystyle{spmpsci}      
\bibliography{references2.bib}   



\end{document}